\documentclass{article}

\usepackage[preprint]{neurips_2024}

\usepackage[utf8]{inputenc} % allow utf-8 input
\usepackage[T1]{fontenc}    % use 8-bit T1 fonts
\usepackage{hyperref}       % hyperlinks
\usepackage{url}            % simple URL typesetting
\usepackage{booktabs}       % professional-quality tables
\usepackage{amsfonts}       % blackboard math symbols
\usepackage{nicefrac}       % compact symbols for 1/2, etc.
\usepackage{microtype}      % microtypography
\usepackage{xcolor}         % colors
\usepackage{graphicx}
\usepackage{amsmath}
\usepackage[table]{xcolor}  % http://ctan.org/pkg/xcolor
\usepackage{enumitem}

\title{The Iconicity of the Generated Image}

\author{%
  Nanne van Noord \\
  University of Amsterdam\\
  \texttt{n.j.e.vannoord@uva.nl} \\
  \And
  Noa Garcia \\
  The University of Osaka\\
  \texttt{noagarcia@ids.osaka-u.ac.jp} \\
}

\begin{document}

\maketitle

\begin{figure}[h!]
\centering
\footnotesize
\newcommand{\figgen}[2]{\includegraphics[width=50pt,height=50pt]{imgs/#1/#2_name.png}&\includegraphics[width=50pt,height=50pt]{imgs/#1/#2_prompt.png}&\includegraphics[width=50pt,height=50pt]{imgs/#1/#2_description.png}}

\begin{tabular}{@{\hspace{0pt}}@{\hspace{2pt}}cc@{\hspace{2pt}}c@{\hspace{2pt}}c@{\hspace{2pt}}cc@{\hspace{2pt}}c@{\hspace{2pt}}c@{\hspace{2pt}}c@{\hspace{0pt}}}

 & & \multicolumn{3}{c}{Stable Diffusion 1.4} & & \multicolumn{3}{c}{Stable Diffusion 3.5} \\
\cline{3-5}
\cline{7-9}\\[-5pt]

iconic image & & \textit{name} & \textit{prompt} & \textit{description} & & \textit{name} & \textit{prompt} & \textit{description} \\[2pt]

\includegraphics[width=50pt,height=50pt]{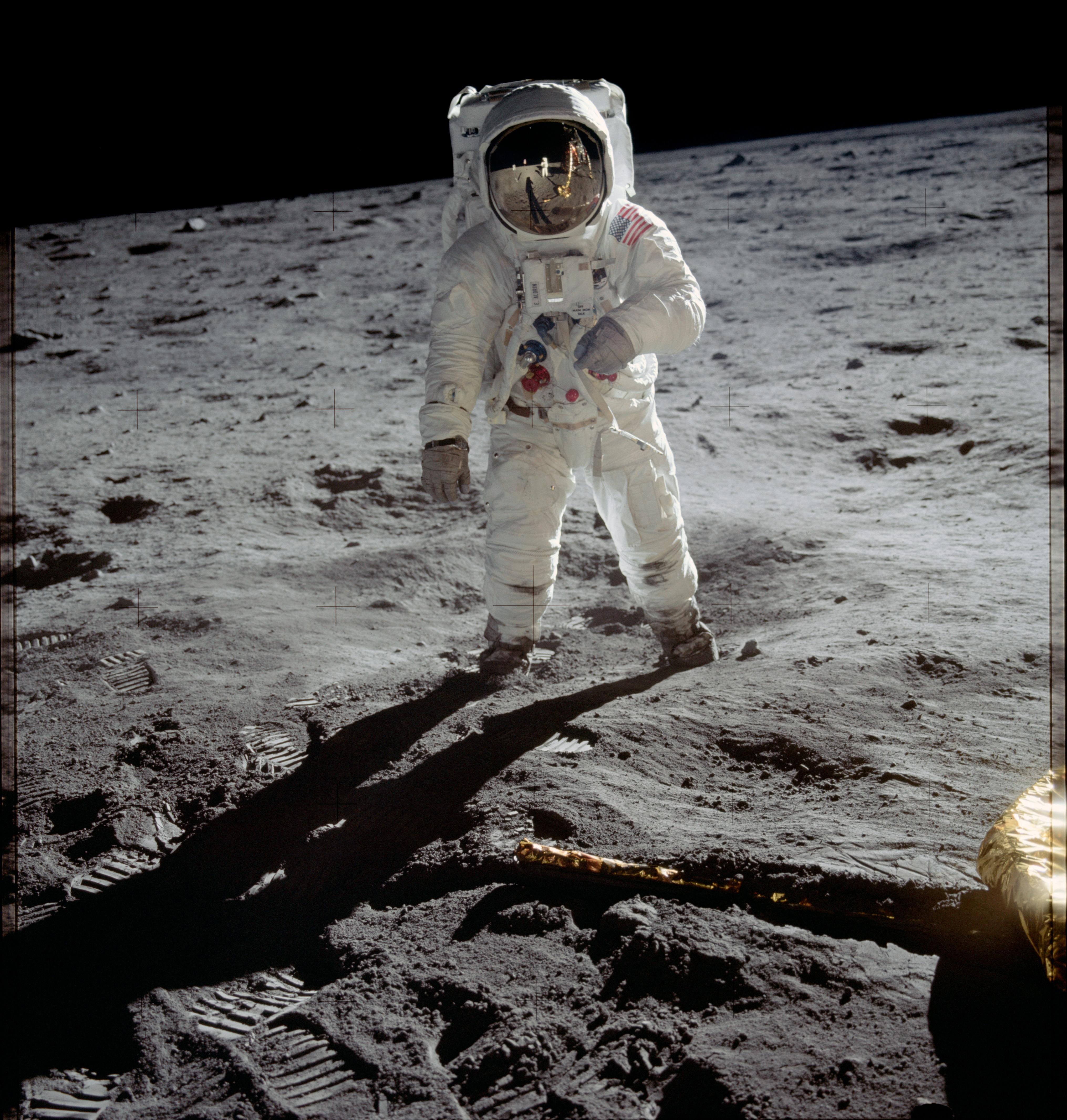} & & \figgen{sd14}{moon}  & & \figgen{sd35}{moon} \\

\includegraphics[width=50pt,height=50pt]{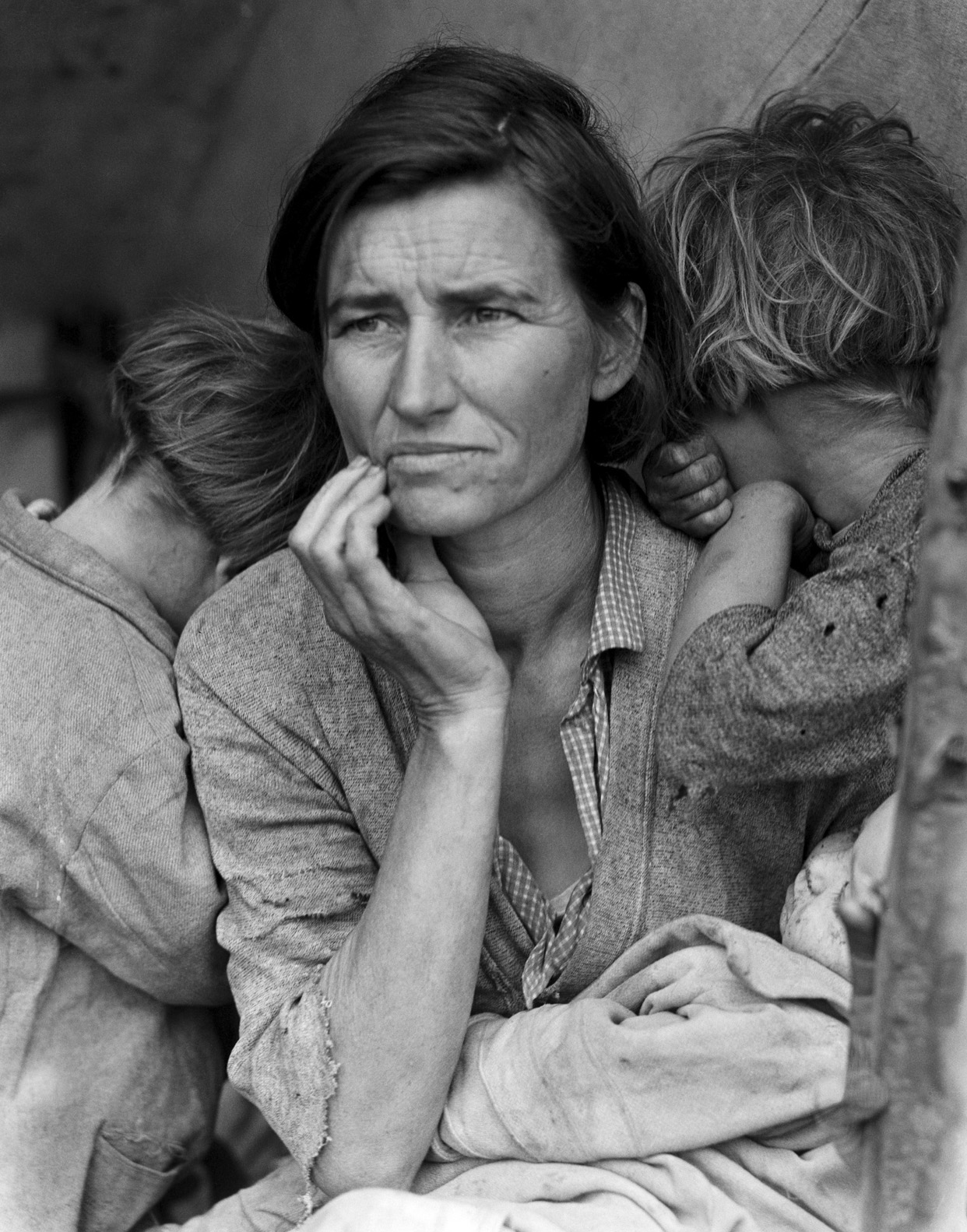} & & \figgen{sd14}{mother} &  & \figgen{sd35}{mother} \\

\includegraphics[width=50pt,height=50pt]{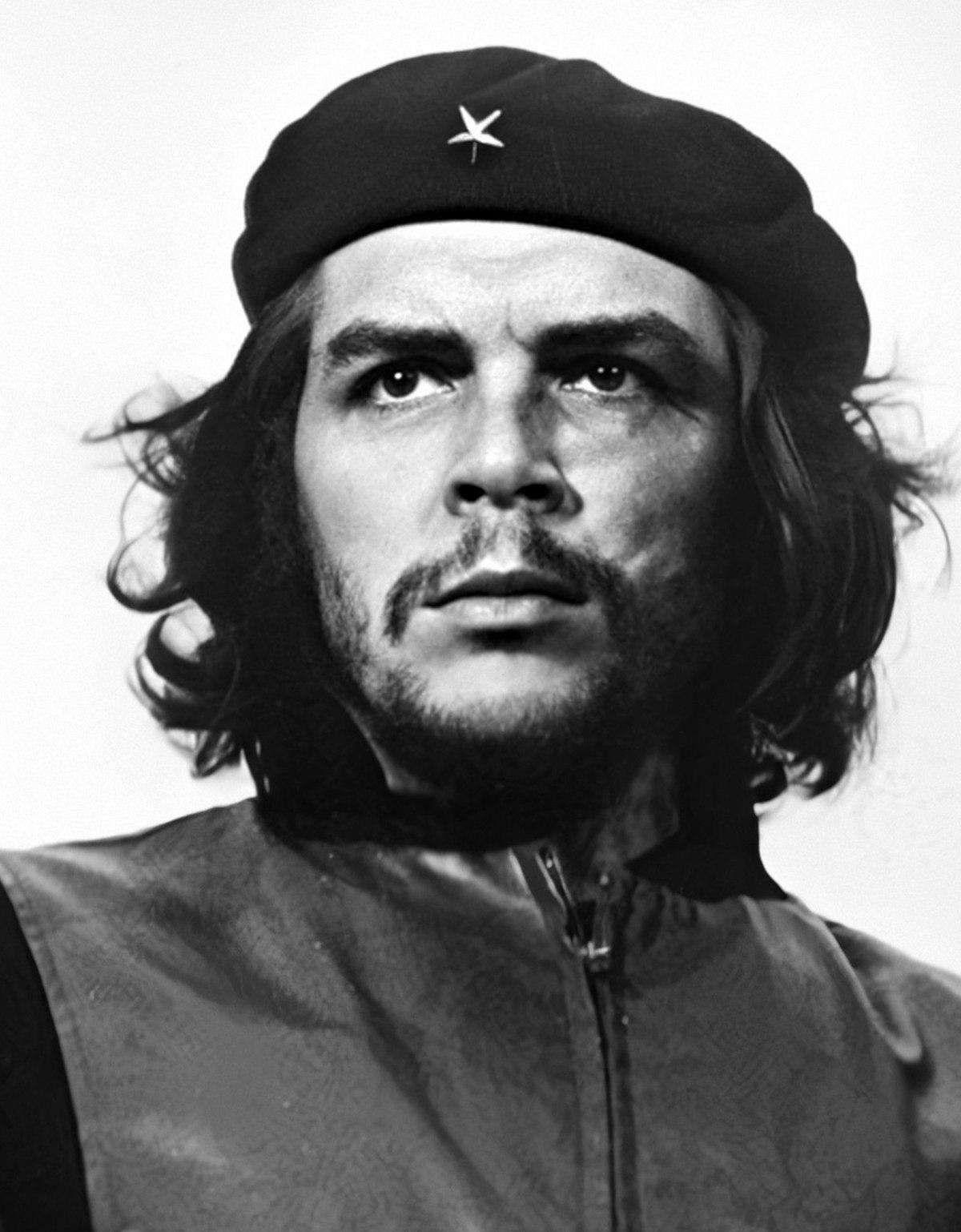} & & \figgen{sd14}{guerrillero} & & \figgen{sd35}{guerrillero} \\

\includegraphics[width=50pt,height=50pt]{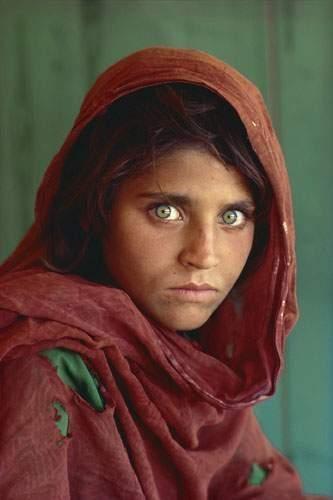} & & \figgen{sd14}{afghan}  & & \figgen{sd35}{afghan} \\[5pt]

\end{tabular}
\caption{\textbf{Iconic images and associated generations}. For each original image (left), we show three generations from Stable Diffusion 1.4 and 3.5 using three prompt types. Different behaviors are observed. While generations for \textit{A man on the moon} (first row) consistently resemble the original, those for \textit{Migrant Mother} (second row) fail to depict white people, exposing stereotype biases associated with the term ``migrant''. \textit{Guerrillero heroico} (third row) only produces a recognizable likeness of Che Guevara when prompted with a description, in contrast to \textit{Afghan girl}, where a descriptive prompt changes the gender of the subject.
 \label{fig:generated-images}}
\end{figure}

\begin{abstract}
How humans interpret and produce images is influenced by the images we have been exposed to. Similarly, visual generative AI models are exposed to many training images and learn to generate new images based on this. Given the importance of iconic images in human visual communication, as they are widely seen, reproduced, and used as inspiration, we may expect that they may similarly have a proportionally large influence within the generative AI process. In this work we explore this question through a three-part analysis, involving data attribution, semantic similarity analysis, and a user-study. Our findings indicate that iconic images do not have an obvious influence on the generative process, and that for many icons it is challenging to reproduce an image which resembles it closely. This highlights an important difference in how humans and visual generative AI models draw on and learn from prior visual communication.
\end{abstract}

\section{Introduction}

The formation of human aesthetic and depictive preferences, represented by how we portray and view objects, scenes, and concepts, is largely influenced by cultural and social factors related to the images we have been exposed to \citep{Palmer_Schloss_Sammartino_2013}. Humans do not interpret images in a vacuum, but within their socio-cultural context, which grants them a wider sense of meaning. For instance, an image that leaves a deep emotional impact on the viewer tends to influence them more than one merely glance upon in passing \citep{Khosla_Raju_Torralba_Oliva_2015}. Visual generative AI models, such as DALL-E \citep{Ramesh_Pavlov_Goh_Gray_Voss_Radford_Chen_Sutskever_2021}, Stable Diffusion \citep{Rombach_Blattmann_Lorenz_Esser_Ommer_2022},  Gemini \citep{saharia2022photorealistic}, or Midjourney \citep{Midjourney2025}, appear to undergo an analogous, though fundamentally distinct, preference-forming process. Through accelerated training, these models are exposed to billions of images scraped from the internet, a process that inherently strips them of their context and reduces them to mere pixels devoid of significance. The training process goal is to adjust the model's weights until it can generate visually appealing content; however, this decontextualized learning represents a fundamental difference to the human experience, making the mechanisms of how influence is incorporated within the generative process of AI models an aspect that remains poorly understood \citep{Wang_Efros_Zhu_Zhang_2023}. 

To investigate how visual knowledge of specific images is integrated into the internal representations of generative AI, one can analyze the behavior of visual generative models in response to images of significant social and cultural influence. The central question is whether these models, trained on decontextualized data, are still influenced by the same kinds of visuals as humans. Iconic images, as representations that capture an historical event, era, or cultural concept into a single frame, are uniquely suited for this analysis. Characterized by their widespread circulation and capacity to evoke strong emotional responses, they function as foundational elements of collective memory, often transcending their original context to acquire broader symbolic meaning \citep{Perlmutter_1998}. Importantly, their widespread online circulation suggests they should be largely represented in training datasets that are, indeed, scrapped from the internet. This, in turn, raises a follow-up question: does online overrepresentation translate into a proportionally larger influence within the model's generative process?

Intuitively, the answer seems to be yes. Given the prominence of iconic images within visual communication, we may expect these same images to be prominent in training datasets, and hence to be influential for generated images -- akin to their influence on human-made images. 
However, common practices like data deduplication or data filtering used to collect visual datasets could distort their representation. 
This problem is compounded by a lack of transparency, as data and model inspection is largely limited to open-source models like Stable Diffusion trained on open-source datasets like LAION-5B \citep{Schuhmann_Beaumont_Vencu_Gordon_Wightman_Cherti_Coombes_Katta_Mullis_Wortsman_et_al_2022}. Investigating the relationship between iconic and generated images is considerably more challenging for opaque and closed-source models. 

Relying on open-source resources, we design a framework to analyze the influence of iconic images on text-to-image generation models, that is, models that create images specifically from a text description or prompt. For an established collection of iconic images from prior work \citep{van_der_Hoeven_2019,Smits_Ros_2020}, we design a set of prompts that facilitate their generation. We then conduct a three-part analysis. In the first part (Section \ref{sec:attribution}), we investigate the contribution of iconic images to generations with data attribution techniques. In the second part (Section \ref{sec:computational_analysis}), we measure how the semantic content of the generated images relates to the original iconic images by computing image similarities in the embedding space. In the final analysis (Section \ref{sec:user_study}), we study the socio-historical representation of the image generations by conducting a user-study. Additionally, through these analyses, we explore the notion of bias in text-to-image models \citep{luccioni2023stable,Wu_Nakashima_Garcia_2024}, to discover whether AI image generators have an \textit{iconicity bias}. 
\section{The Iconic Image}
\label{sec:iconic_image}
As generated images become increasingly prevalent, there is a potential for them to reshape how and what we remember about historical events \citep{Stokel-Walker_2023,Sheng_Tuyttens_Keyserlingk_2025}. Iconic images, those highly recognized photographs that visually represent historical events as well as culturally and societally prominent concepts, have played a central role in shaping this memory within traditional media like history books \citep{Kleppe_2013} and journalism \citep{Dahmen_Mielczarek_Perlmutter_2018}. While these images have been extensively studied in those fields \citep{Perlmutter_1998, Lucaites_Hariman_2001, Dahmen_Miller_2012}, limited work has considered them computationally \citep{Berg_Berg_2009, van_Noord_2022} and their influence within the emerging domain of AI-generated imagery remains unexplored. As far as we are aware, ours is the first work to study them in the context of image generation.

Iconic images function as cultural memories \citep{Cohen_Boudana_Frosh_2018}, entering and persisting in the collective consciousness through continuous circulation, reproduction, and reinterpretation. A key criterion for their iconicity is their frequency and breadth of dissemination \citep{Perlmutter_1998,Dahmen_Mielczarek_Perlmutter_2018}. It is thus unavoidable that these widely reproduced images are present in the large-scale, web-scraped datasets used to train text-to-image models.\footnote{Empirically we verified that all $26$ iconic images studied are represented in the LAION-5B dataset.} Furthermore, the phenomenon of intericonicity, where new images reference the composition or subject matter of prior icons, reinforces their status and amplifies their influence on visual culture \citep{van_Noord_2022}. 

Within iconic images, we may distinguish between two categories.  \cite{Perlmutter_1998} introduces the notion of \textit{discrete} iconic images, which are singular hyper-famous images that are tied to specific events, and \textit{generic} iconic images, which describe a common theme through recurring situations and shared characteristics. Within the former category we may find images such as Albert Einstein sticking out his tongue, or the tragic scene of children escaping the aftermath of a napalm attack during the Vietnam war. The latter category does not involve specific images, but rather taps into common depictions of recurring scenes, for example, those of an emaciated African child or a polar bear on melting ice \citep{van_Noord_2022}. Given their defined singular nature, discrete iconic images provide a more tractable starting point for computational analysis. While this work focuses exclusively on discrete iconic images, studying the influence of generic iconic imagery on image generation is an important avenue for future research directions.

% ######## TABLE
\begin{table}[t]
\footnotesize
\centering
\caption{\textbf{List of iconic images.} The $26$ iconic images by name and photographer together with their \textit{description} prompt. \textit{Description} prompts start with ``Photograph of'', trimmed for formatting.}
\label{tab:iconic_photos}
\begin{tabular}{p{3cm}p{2.5cm}cp{6cm}}
\toprule
\textbf{Name} & \textbf{Photographer} & \textbf{Year} & \textbf{Description prompt} \\
\midrule
Migrant Mother & Dorothea Lange & 1936 & destitute mother with her children \\
Falling soldier & Robert Capa & 1936 & a soldier being shot during the Spanish civil war \\
Hindenburg disaster & Sam Shere & 1937 & the Hindenburg zeppelin descending in flames \\
Raising the flag on Iwo Jima & Joe Rosenthal & 1945 & marines raising a flag during the battle of Iwo Jima \\
Raising a flag over the Reichstag & Yevgeny Khaldei & 1945 & a Soviet soldier raising a flag over the Reichstag during World War II \\
V-J Day in Times Square & Alfred Eisenstaedt & 1945 & a male sailor kissing a female dental assistant against her will in 1945 \\
Holocaust survivors & Lee Miller & 1945 & the former prisoners of the "little camp" in Buchenwald stare out from the wooden bunks in which they slept three to a "bed" \\
Gandhi and the spinning wheel & Margaret Bourke-White & 1946 & Mahatma Gandhi spinning khadi yarn with a wheel \\
Founding of the PRC & Hou Bo & 1949 & proclamation of the People's Republic of China by Mao Zedong \\
Guerrillero heroico & Alberto Korda & 1960 & Argentine revolutionary Che Guevara \\
Assassination of Inejiro Asanuma & Yasushi Nagao & 1960 & a man stabbing Inejiro Asanuma \\
Burning monk & Malcolm Browne & 1963 & the monk Quang Duc during his self-immolation \\
Saigon execution & Eddie Adams & 1968 & the execution of a Viet Cong soldier in Saigon during the Vietnam war \\
A man on the moon & Neil Armstrong & 1969 & astronaut Buzz Aldrin on the moon \\
Kent State shootings & John Filo & 1970 & the killing of four and wounding of nine unarmed college students by the Ohio National Guard on the Kent State University campus \\
Accidental napalm & Nick Ut & 1972 & a young girl running on a road after being severely burned by a napalm attack \\
Allende's last stand & Luis Orlando Lagos & 1973 & the last day of President Allende during the 1973 Chilean coup d'état \\
Afghan girl & Steve McCurry & 1984 & an Afghan refugee in Pakistan during the Soviet–Afghan War \\
Tank man & Jeff Widener & 1989 & a man stood in front of a column of tanks leaving Tiananmen Square \\
Vulture and the little girl & Kevin Carter & 1993 & a frail famine-stricken child collapsed in the foreground with a hooded vulture eyeing from nearby \\
Survivor of a Hutu death camp & James Nachtwey & 1994 & a survivor of a Hutu death camp who was attacked with machetes during the Rwanda Genocide \\
Falling man & Richard Drew & 2001 & a man falling from the World Trade Center during the September 11 attacks in New York City \\
Hijacked airplane & unknown & 2001 & a hijacked plane on 911 hitting the twin towers \\
Abu Ghraib prisoner & Ivan Frederick & 2003 & a prisoner at Abu Ghraib prison with wires attached to his fingers, standing on a box with a covered head \\
Situation room & Pete Souza & 2011 & U.S. president Barack Obama and his national security team in the White House Situation Room \\
Alan Kurdi & Nilüfer Demir & 2015 & a two-year-old boy lying dead on the beach \\
\bottomrule
\end{tabular}
\end{table}
% ######## TABLE

To guide our analysis, we use a collection of $26$ discrete iconic images used in prior work \citep{Smits_Ros_2020}. Although there is variation in terms of year and region within the collection, it predominantly consists of images from the mid to late 1900s from a US-centric perspective. Whilst there is no definitive list of iconic images, \cite{Smits_Ros_2020} retrieved a total of $940$ thousand online circulations, confirming that the selected images are widely available on the web and frequently redistributed -- a clear hallmark of iconicity. The full list of the iconic images by name and photographer is provided in Table \ref{tab:iconic_photos}.
\section{The Generated Image}
\label{sec:generated_image}

Within the field of generative AI, text-to-image models have demonstrated incredible potential for translating human produced textual descriptions to realistic images \citep{Zhang_Zhang_Zhang_Kweon_Kim_2024}. Early works involved text-conditional GAN models \citep{Reed_Akata_Yan_Logeswaran_Schiele_Lee_2016} as the first attempt to go directly from text to images. These initial results were improved upon with autoregressive models \citep{Ramesh_Pavlov_Goh_Gray_Voss_Radford_Chen_Sutskever_2021}, but the key breakthrough in the widespread adoption of text-to-image models has been driven by diffusion models, such as Stable Diffusion \citep{Rombach_Blattmann_Lorenz_Esser_Ommer_2022}. Due to these technological innovations, the quality of generated images has gone from unrealistic and pixelated to photorealistic and imaginative. The comprehensive survey by \cite{Zhang_Zhang_Zhang_Kweon_Kim_2024} describes this development and the role of diffusion models in detail.

A growing body of research has moved beyond mapping the technological development of text-to-image models to critically examine how they function, and in particular, what their limitations and biases are \citep{bianchi2023easily,luccioni2023stable,Wu_Nakashima_Garcia_2024, foka2024framework, Sheng_Tuyttens_Keyserlingk_2025, Shaw_Ye_Krishna_Zhang_2025, De_Rosa_Palmini_Cetinic_2025}. These studies have revealed that text-to-image models frequently reproduce stereotypes \citep{bianchi2023easily}, amplify societal biases like gender \citep{luccioni2023stable, Wu_Nakashima_Garcia_2024}, romantize complex industries like livestock farming \citep{Sheng_Tuyttens_Keyserlingk_2025}, and contribute to a homogenization of visual output \citep{De_Rosa_Palmini_Cetinic_2025}. A key driver of these issues is the models' training on large-scale, web-scraped datasets, which reflect the statistical imbalances of online visual culture \citep{birhane2021multimodal, garcia2023uncurated}.  Yet, the influence of widespread cultural artifacts, such as  iconic images, remains poorly understood. 

To generate images,
we rely on the most popular open-source family of models: Stable Diffusion (SD) \citep{Rombach_Blattmann_Lorenz_Esser_Ommer_2022}. 
We employ two versions, SD1.4 and SD3.5, for generalization and  comparative analysis. Whilst SD3.5 is newer and has improved visual fidelity, it employs different training strategies from its predecessor. Running both models enables a direct comparison of how these differences, as well as gains in visual quality, impact the generated output.

\paragraph{Prompt design} For each of the $26$ iconic images in Section \ref{sec:iconic_image}, we generate three types of prompts: (1) \textit{name}, which uses the canonical name of the iconic image as in \cite{Smits_Ros_2020}; (2) \textit{prompt}, which places the name into the template ``A photograph of \textit{name}''; and (3) \textit{description}, which is a sentence describing the setting and content of the original iconic image. The full list of prompts is provided in Table~\ref{tab:iconic_photos}.

\paragraph{Generation details} For SD1.4 we follow the configuration as described in \cite{Wang_Efros_Zhu_Zhang_2023}, with a guidance scale of $6$. For SD3.5 we perform $20$ inference steps with a guidance scale of $7$, and the negative prompt ``lowres, low quality, worst quality''. All images were generated sequentially for each prompt without manual intervention.
For each iconic image and prompt type, we generate $50$ images using both SD1.4 and SD3.5, resulting in a corpus of $7,800$ generated images ($26$ iconic images $\times$ $3$ prompts $\times$ $50$ samples $\times$ $2$ models). A few  examples are shown in Figure \ref{fig:generated-images}.

\section{Data Attribution Study}
\label{sec:attribution}

Data attribution is a technique that aims to quantify the influence of the training data on a generated image \citep{Wang_Efros_Zhu_Zhang_2023, Wang_Hertzmann_Efros_Zhu_Zhang_2024, Lin_Tao_Dong_Xu_2024}. A naive solution to data attribution consists on retraining the generation model without a given training image and check how the output differs. However, retraining a generative model for each dataset permutation where a single image is removed is unfeasible. Instead, alternatives approximate this behavior by using specific predictive models trained to detect attribution influence. For example,  \cite{Wang_Efros_Zhu_Zhang_2023} trains a data attribution detector with synthetic data by generating variants of an image, which results in known attributions. Alternatively, \cite{Wang_Hertzmann_Efros_Zhu_Zhang_2024} propose to force a network to ``unlearn'' a particular generated image, which results in the model unlearning the training images that are necessary to generate that image, which is used to detect the influential training images. Moving beyond synthetic settings, we apply data attribution detectors to determine whether iconic images in the training set have a greater influence on all generated images than non-iconic images. 

\paragraph{Setup} To quantify whether the influence of iconic images can be determined for generated images, we use the data attribution detector by \cite{Wang_Efros_Zhu_Zhang_2023}. This approach uses an evaluation dataset of 1M images that consists of images randomly sampled from LAION-400M  \citep{Schuhmann_Vencu_Beaumont_Kaczmarczyk_Mullis_Katta_Coombes_Jitsev_Komatsuzaki_2021} and from a dataset of images which are generated conditioned on a known image and text prompt. This latter set of generated images allows for attribution training, as the image on which the synthetic image is conditioned is known. We follow a similar setup in that we use the same 1M subset, to which we add the $26$ iconic images ensuring that we have a known instance of each iconic image present for automatic evaluation. These 1M+26 images form the database against which we perform the attribution.

\begin{figure}[h]
    \centering
    \includegraphics[width=0.7\linewidth]{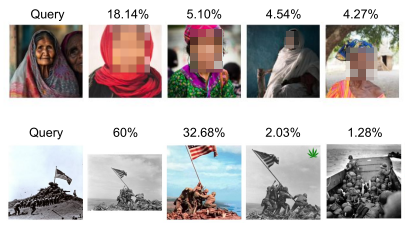}
    \caption{\textbf{Data attribution examples}. For two query images generated with the prompts `\textit{Photograph of the migrant mother}' and `\textit{Photograph of raising the flag on Iwo Jima}', respectively, we show the top-4 images in the 1M+26 database with highest attribution score.}
    \label{fig:attribution_ex}
\end{figure}

To perform the attribution, each generated image $x_i \in X$ is passed through the attribution detector $f$, resulting in an influence score vector $s_i = f(x_i)$, where $s_i \in \mathbb{R}^N$ with $N$ being the size of the attribution database. Concretely, for each database image an attribution scores is obtained, represented by a calibrated probability $0-100$, where higher values indicate greater probability that the generated image was influenced by that database image. The top influence scores for two queries are shown in Figure~\ref{fig:attribution_ex}.

Using the 1M+26 dataset, we conduct an attribution experiment where for each prompt (across all three prompt types) we perform the attribution for the $50$ images generated with SD3.5. Subsequently, the attribution scores are averaged over the $50$ images, to reduce variation due to generated image differences, resulting in an attribution ranking for all $26$ iconic images across the three prompt types. To determine whether an iconic image has influenced the generation, we check whether it is present in the top of the attribution ranking, where we use $1\%$ influence probability as the cut-off. As such, any iconic image which is considered influential by the data attribution method for a certain prompt will be identified.

\paragraph{Results}
Results are reported in Table~\ref{tab:attribution}. We can observe that most iconic images ($18$ out of $26$) are found to be influential for the \textit{description} type of prompt, whereas for the other two types only $8$ are influential. Moreover, we find that for certain prompts the iconic image is consistently attributed across the prompts (e.g., \textit{Raising the flag on Iwo Jima}), whereas for other prompts (e.g., \textit{Survivor of a Hutu death camp}) the iconic image is never found. In terms of the influence score, we do not see a clear trend, however, often times the iconic image is assigned a low influence score, despite the prompts being tailored to generate that image. 

\begin{figure}[h]
    \centering
    \includegraphics[width=0.6\linewidth]{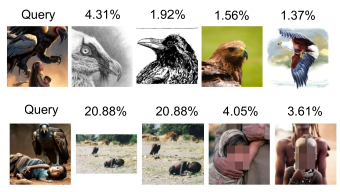}
    \caption{\textbf{Difference in attribution per prompt type}. The query in the first row was generated with the prompt `\textit{Vulture and the little girl}', while the query in the second row was generated from `\textit{Photograph of a frail famine-stricken child collapsed in the foreground with a hooded vulture eyeing from nearby}', both intended to target the same iconic image. Whilst the images generated for the former prompt are not attributed to the iconic image, those by the latter prompt are. The iconic image appears twice as the database contains our added version, and a version that was  present in LAION.}
    \label{fig:attribution_res}
\end{figure}

Figure~\ref{fig:attribution_res} shows how the iconic image title does not result in attribution, whereas a very descriptive prompt does. Notably, this attribution appears to depend primarily on semantic similarity, as the composition and aesthetics do not match the iconic image.

Based on these results, it appears challenging to find the influence of iconic image through data attribution, even when generating images based on prompts that directly refer to the iconic image. Variation between iconic images seems to stem from how closely the generation matches the iconic image. We  explore this further in the next section by quantifying the degree of similarity.

\begin{table}[t]
\footnotesize
\centering
\caption{\textbf{Data attribution results.} Results of data attribution with the influence score (0-100) of the iconic image, in case of multiple instances of the iconic image being retrieved only the score of the top image is reported. Influence scores below the threshold (1\%) are indicated with a dash.}
\label{tab:attribution}
\begin{tabular}{lccc}
\toprule
Iconic image & \textit{name} & \textit{prompt} & \textit{description} \\
\midrule
Migrant Mother & - & - & 2.8 \\
Falling soldier & - & - & - \\
Hindenburg disaster & 14.6 & 8.0 & 32.8 \\
Raising the flag on Iwo Jima & 32.6 & 60.0 & 50.7 \\
Raising a flag over the Reichstag & - & - & 6.4 \\
V-J Day in Times Square & 1.2 & - & 2.3 \\
Holocaust survivors & - & - & 3.7 \\
Gandhi and the spinning wheel & 1.7 & 17.2 & 15.2 \\
Founding of the PRC & - & - & - \\
Guerrillero heroico & - & - & 69.4 \\
Assassination of Inejiro Asanuma & 5.1 & - & 7.1 \\
Burning monk & - & - & - \\
Saigon execution & - & - & - \\
A man on the moon & 9.0 & 9.0 & 12.6 \\
Kent State shootings & 2.5 & 1.9 & - \\
Accidental napalm & - & 2.0 & 1.3 \\
Allende's last stand & - & - & 1.5 \\
Afghan girl & 14.2 & 14.6 & 1.1 \\
Tank man & - & 1.3 & 3.0 \\
Vulture and the little girl & - & - & 20.9 \\
Survivor of a Hutu death camp & - & - & - \\
Falling man & - & - & 5.1 \\
Hijacked airplane & - & - & 32.7 \\
Abu Graib prisoner & - & - & - \\
Situation room & - & - & 1.1 \\
Alan Kurdi & - & - & - \\
\bottomrule
Count & 8 & 8 & 18 \\
\bottomrule
\end{tabular}
\end{table}
\section{Computational Similarity Analyses}
\label{sec:computational_analysis}

Next, we computationally evaluate the similarity between the generated images and the original iconic images. The goal is to verify whether prompts engineered around an iconic image can produce images that resemble it, with a high similarity indicating the generative model has some internal knowledge about the original iconic image. We conduct two analyses: icon similarity (Section \ref{sec:icon_similarity}) and image variation (Section \ref{sec:image_variation}). Icon similarity represents how closely the iconic image is matched to the generated images given a particular prompt, while image variation measures the semantic similarity among the generated images, indicating their degree of heterogeneity. When an iconic image is very influential on the generation process, we would see  high icon similarity and low image variation. Whereas if the iconic image is unfamiliar to the model, we would see low icon similarity.

To contextualise our results, we compare generated images' similarity to that achieved by non-generated data sources. In particular, we collect image data to reference against from two sources: Google Image Search\footnote{\url{https://images.google.com/}, accessed: 21st March 2025.} (GIS) and Have I Been Trained\footnote{\url{http://www.haveibeentrained.com}, accessed: 21st March 2025.} (HIBT). The GIS set is based on webdata shaped through the lens of the underlying search algorithms, as such it gives insight into how tightly a certain prompt is tied to visual concepts. The HIBT tool enables text-based search through the five billion images in LAION-5B \citep{Schuhmann_Beaumont_Vencu_Gordon_Wightman_Cherti_Coombes_Katta_Mullis_Wortsman_et_al_2022}. The retrieved HIBT images reflect how iconic images are represented in the training data of text-to-image models for the provided queries. For all $26$ iconic images and $3$ prompt types, we collect $50$ images from GIS and $50$ images from HIBT, resulting in two reference datasets, totaling $7,800$ images. 

We encode all images ($26$ iconic, $7,800$ generated, and $7,800$ reference) into vector representations using a pre-trained CLIP image encoder \citep{radford2021learning}. CLIP converts both images and text into a shared embedding space, placing semantically related content closer together. This allows to compute how similar two vectors in the semantic space are by simply computing their cosine similarity. Images with high cosine similarity in the embedding space are expected to share highly similar semantic content. 

Given these properties, we propose the \texttt{icon2group} metric to quantify icon similarity as
\begin{equation}
    \texttt{icon2group} = \frac{1}{N} \sum_{j} \text{cos}(I, G_j),
\end{equation}
and the \texttt{ingroup} metric to measure image variation as
\begin{equation}
    \texttt{ingroup} = \frac{1}{N(N-1)} \sum_{j \neq k} \text{cos}(G_j, G_k),
\end{equation}
where $I$ is the CLIP embedding of the iconic image, $G_j$ and $G_k$ are the CLIP embeddings of the $j$-th and $k$-th generated images, respectively, $\text{cos}(\cdot, \cdot)$ is the cosine similarity between two embeddings, and $N$ is the number of generated images per icon and prompt type.

\subsection{Icon Similarity}
\label{sec:icon_similarity}

\begin{table}[t]
\centering
\renewcommand{\arraystretch}{1.1}
\footnotesize
\caption{\textbf{Icon similarity results.} Results for the \texttt{icon2group} metric per type of prompt.}
\label{tab:iconsim}
\begin{tabular}{lp{2cm}p{1.5cm}p{1.5cm}p{1.5cm}}
\toprule
& Data source & \textit{name} &  \textit{prompt} &  \textit{description} \\
\midrule

Generated data & SD1.4 &  $0.643$ &  $0.692$ &	 $0.745$\\
& SD3.5 &  $0.652$ &  $0.678$ &	 $0.735$\\
\midrule
Reference data & GIS &  $0.762$ &  $0.794$ &  $0.754$ \\
& HIBT &  $0.670$ &  $0.538$ &  $0.165$\\

\bottomrule
\end{tabular}
\end{table}

The results for the icon similarity analysis are presented in Table~\ref{tab:iconsim}. For the two Stable Diffusion models (SD1.4 and SD3.5) we can observe that as the prompt becomes more detailed and specific, from \textit{name} to \textit{description}, there is an increase in similarity, with the \textit{description} prompts leading to the highest semantic similarity between the generated and iconic images. This indicates that the engineered descriptions are effective, as they  generate images more semantically aligned with the iconic image than using the name alone.

When comparing the two versions of Stable Diffusion, SD1.4 achieves a slightly higher similarity for the more detailed prompts, but overall their scores are very similar. This highlights that the semantic similarity metric is not sensitive to the visual fidelity, as the SD3.5 results are of much higher visual fidelity, but instead focus on the semantics.

For the reference data, we observe different trends. The HIBT data, retrieved from the LAION-5B dataset, show a strong degradation in similarity as the prompts become more detailed. This is because, for a number of the \textit{description} prompts, the HIBT tool returned no results, which gives an \texttt{icon2group} similarity of $0$ for that particular prompt and lowers the average metric to $0.165$. We suspect this is caused by the HIBT's retrieval system based on text-to-text similarity, which may be more sensitive to exact keyword matches between the query (in our case the prompt) and the dataset image's captions, as opposed to semantic matches. 

The GIS data obtained via Google Search, on the other hand, shows very little influence from the prompt type, with it obtaining the highest \texttt{icon2group} similarity across all three prompt types. This confirms the notion that these prompts are strongly associated with web-published images. Moreover, when using GIS as a reference we can observe that the generated images for the \textit{description} prompt obtain nearly the same \texttt{icon2group} similarity, thereby highlighting that the generated images are highly similar to the iconic images.

\subsection{Image Variation}
\label{sec:image_variation}
The results for the image variation analysis are presented in Table~\ref{tab:ingroup}. A notable observation  is that the prompt type has little influence on the \texttt{ingroup} metric, with only HITB results showing a reduction in similarity, again probably caused by a failure in the text-based retrieval system. The GIS \texttt{ingroup} similarity is moderately high, which relates to the search results containing many instances of the iconic image, but also strongly associated images such as related photographs, book covers, or photographs of the photographer themselves.

The generated data has, in general, higher \texttt{ingroup} similarity scores, which means there is little semantic variation amongst them.
When comparing SD1.4 and SD3.5 there are minor differences, with SD3.5 having a slightly higher \texttt{ingroup} similarity. This may be due to the guidance scale parameter used, which was $6$ for SD1.4 and $7$ for SD3.5. Whilst these values are not directly comparable, both models were guided towards strong adherence to the prompt to maximally determine how well they can generate images resembling the iconic images. 

\begin{table}[t]
\centering
\renewcommand{\arraystretch}{1.1}
\footnotesize
\caption{\textbf{Image variation results.} Results ofor the \texttt{ingroup} metric per type of prompt.}
\label{tab:ingroup}
\begin{tabular}{lp{2cm}p{1.5cm}p{1.5cm}p{1.5cm}}
\toprule
& Data source & \textit{name} &  \textit{prompt} &  \textit{description} \\
\midrule

Generated data & SD1.4 &  $0.78$ &  $0.81$ &	 $0.84$\\
& SD3.5 &  $0.85$ &  $0.87$ &	 $0.88$\\
\midrule
Reference data & GIS &  $0.71$ &  $0.73$ &  $0.69$ \\
& HIBT &  $0.65$ &  $0.52$ &  $0.17$\\

\bottomrule
\end{tabular}
\end{table}
\section{Socio-Historical Analyses} 
\label{sec:user_study}

While the analyses in the previous sections aimed to identify patterns of influence in the generated images through data attribution and semantic similarity, a complete picture of iconicity requires studying its socio-historical context. The goal is to identify whether the generated images are correctly situated in their original socio-historical context and whether they exhibit any  emerging biases. Given that knowing the connection between the generated and the original images could result in confirmation bias, for the socio-historical analyses we rely on a user study with volunteers who had no prior exposure to the generated images.

The user study consists of a two part survey. In the first survey, we ask questions about the generated images alone, and whether the volunteers can relate them to any image they have previously seen before. In the second survey, we compare the generated images against the originals. The study is conducted with eight participants ($3$ women, $5$ men, age $25-30$) from different countries ($4$ China, $2$ Netherlands, $2$ Philippines-US) each evaluating $52$ generated images. Images are generated with the \textit{name} and \textit{description} prompts for each of the $26$ iconic images. From the pool of images generated by Stable Diffusion 3.5, we randomly select one generation per iconic image and prompt type.

\subsection{Generated image survey}
In the first survey, participants are shown a generated image and asked questions about its historicity, theme, and recognizably. The goal is to assess whether the AI-generated images retain key elements that make them culturally familiar and convey meaningful cues about the source image's historical moment. The questions and their answers are:

\subsubsection*{Historicity}

\textit{Question.} Which historical context do you think this image represents?

\textit{Answer.} We provide participants with a list of five historical events and ask them to select the one that best matches the generated image. Among these, one event corresponds to the true context of the iconic image, while the other four are generated using the multimodal language model Qwen2.5-VL-32B-Instruct~\citep{Qwen2.5-VL}. Specifically, given a generated image as input, the language model is asked to produce two plausible but incorrect historical contexts and two unrelated contexts that are clearly mismatches. The proposed contexts are manually curated to ensure quality. 

\begin{figure}
    \centering
    \includegraphics[width=0.9\linewidth]{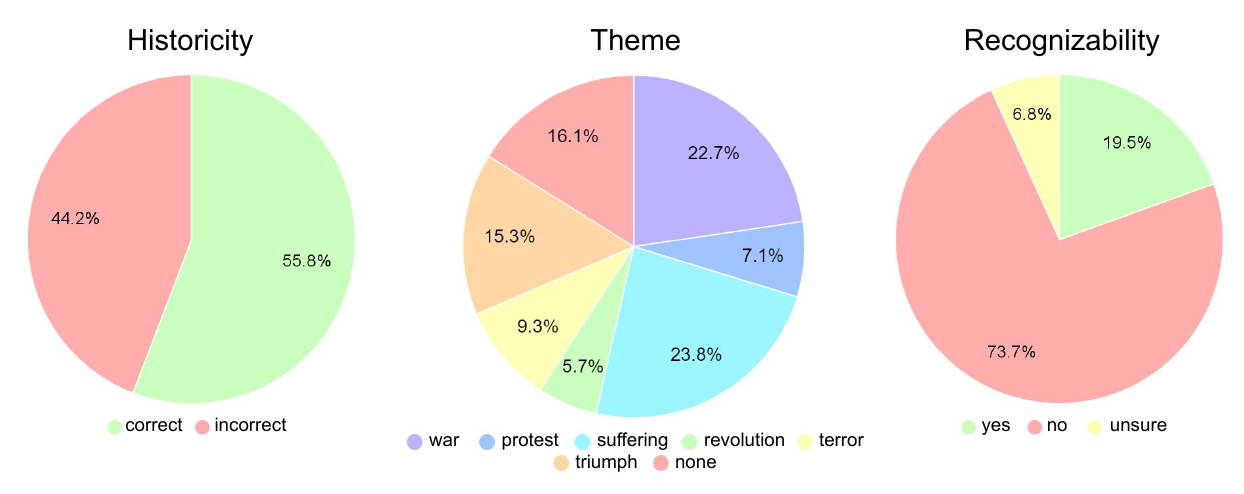}
    \caption{\textbf{Generated image survey results}. Each pie chart shows the results for each question: historicity (left), theme (middle), and recognizability (right).}
    \label{fig:user-study-1}
\end{figure}

\subsubsection*{Theme}

\textit{Question.} Which theme do you think this image represents?

\textit{Answer.} With assistance from ChatGPT, we curate a list of six themes, \textit{War}, \textit{Protest}, \textit{Suffering}, \textit{Revolution}, \textit{Terror}, and \textit{Triumph}, derived from the $26$ original iconic images. Participants must select the most fitting theme or choose \textit{None} if the image does not align with the provided options. 

\subsubsection*{Recognizability}

\textit{Question.} Do you know the image that this generated image is based on?

\textit{Answer.} Provided options are `\textit{yes}', `\textit{no}', and `\textit{unsure}'. If the answer is affirmative, participants are prompted to either name the original image or describe it in their own words.

\paragraph{Results} The overall results are presented in Figure \ref{fig:user-study-1}. The historical context is correctly identified only in $55.8\%$ of cases (with random chance being $20\%$), indicating that the image generation process often fails to situate the generated images in the intended historical context. Results for theme show responses distributed across categories, with a preference for \textit{suffering} (23.8\%) and \textit{war} (22.7\%). The recognizability of the generated images is notably low, with approximately three out of every four images not recognized by our volunteers. Moreover, in four cases, a participant incorrectly recognized an image as inferred from their descriptions, however, this only happened for images with low recognizability overall. There appears to be no relationship between the year of the photograph and its recognizability.

\begin{figure}[h]
    \centering
    \includegraphics[width=0.85\linewidth]{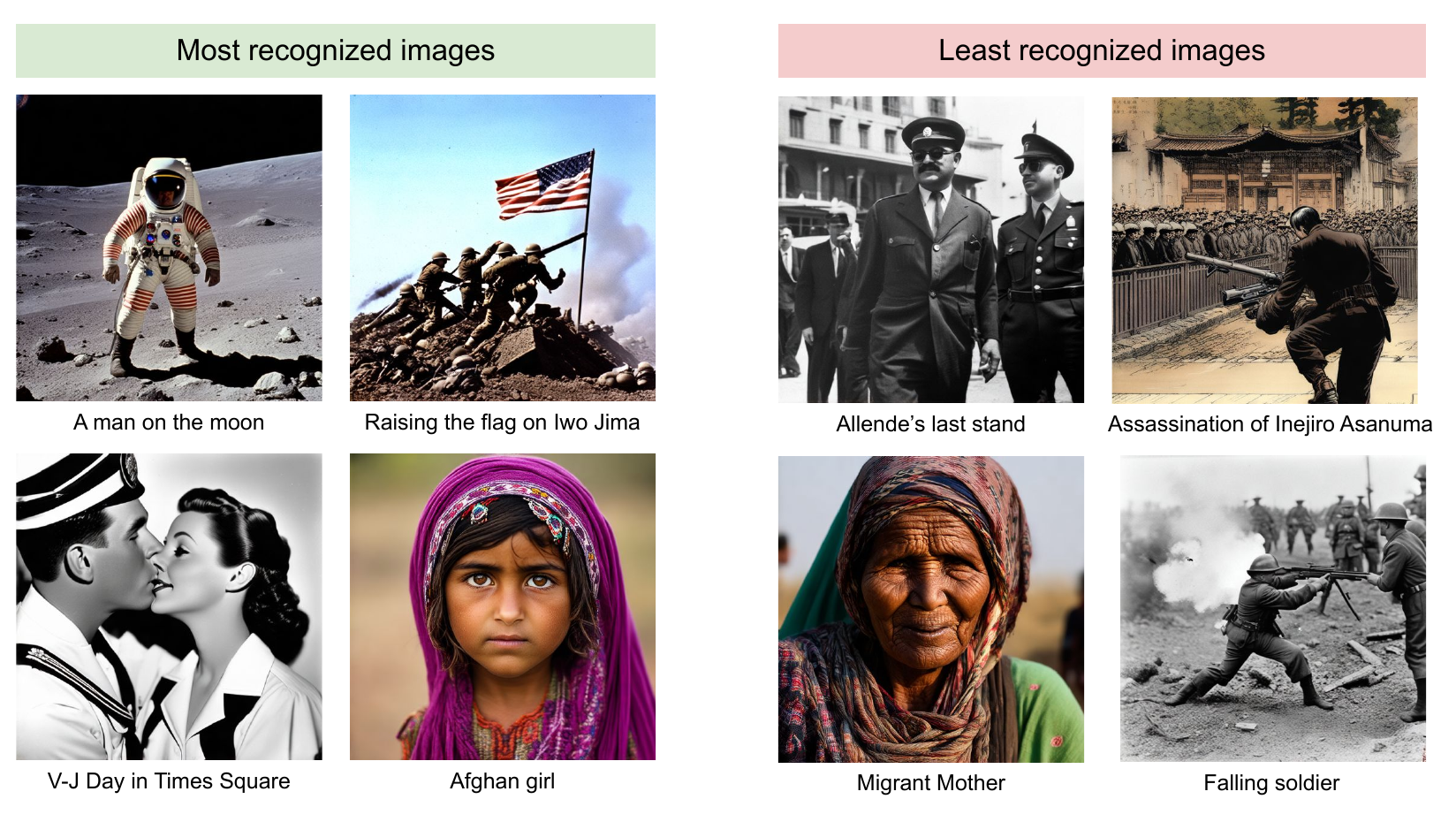}
    \vspace{-10pt}
    \caption{\textbf{Most and least recognized generated images}. The most recognized generations (left) retain key elements of the original iconic photos, while the least recognized ones (right) incorporate severe alternations to their aesthetics or main subject demographics. }
    \label{fig:recog-samples}
\end{figure}

A detailed per-image analysis and prompt type is provided in Table \ref{tab:userstudy}. The use of detailed prompts (\textit{description}) substantially improves historicity, with notable exceptions like \textit{Saigon execution} generated images. The most dramatic improvement is for \textit{Guerrillero heroico}, where historical correctness raises from $0$ to $100\%$. Theme identification is considerable consistent across prompt types, with $17$ out of $26$ iconic images having the same most frequent theme for both prompts. Recognizability also benefits from detailed prompts, doubling the total rate from $13\%$ (with \textit{name} prompts) to $26\%$ (with \textit{description} prompts). The most recognized images are \textit{A man on the moon} and \textit{V-J Day in Times Square}, while six images are not recognized by any participant with any prompt. This is because, as shown in Figure \ref{fig:recog-samples}, the most recognized images contain identifiable elements from their iconic originals, unlike those that went unrecognized.

\begin{table}[t]
\footnotesize
\centering
\caption{\textbf{Generated images survey results per-image and prompt type}. Historicity is reported as the percentage of correct answers, with results higher than 80\%  highlighted in \textbf{bold}. Theme shows the most frequent answer. Recognizability is the percentage of self-reported recognized images, with results higher than 50\% highlighted in \textbf{bold}.}
\label{tab:userstudy}
\resizebox{\textwidth}{!}{\begin{tabular}{lcrrcrrcrr}
\toprule
& & \multicolumn{2}{c}{\textbf{Historicity}} & & \multicolumn{2}{c}{\textbf{Theme}} & & \multicolumn{2}{c}{\textbf{Recognizability}} \\
\cline{3-4}
\cline{6-7}
\cline{9-10}
\textbf{Iconic image} & & \textit{name} & \textit{description} & & \textit{name} & \textit{description} & & \textit{name} & \textit{description} \\
\midrule
Migrant Mother & & 0 & 0 & & suffering & suffering & & 0 & 0 \\
Falling soldier & & 0 & 14 & &	war & war & & 0 & 0 \\
Hindenburg disaster & & \textbf{100} & \textbf{86} & & suffering & terror & &	17 & 14 \\
Raising the flag on Iwo Jima & & 71 & 71 & &  war &  war	& & 43 & 43 \\
Raising a flag over the Reichstag & & 50 & 43 & &  war &  war	& & 0 &	0
\\
V-J Day in Times Square & & 71 & \textbf{100} & &	 triumph &  triumph & & 0	& \textbf{83} \\
Holocaust survivors & & 57 & \textbf{86}	& &  suffering &  suffering & & 0 &	29 \\
Gandhi and the spinning wheel & & 43 & \textbf{86} & &  none &  none & & 14 & 14 \\
Founding of the PRC & & 50 & 14 & &	 triumph &  triumph & & 17 &	0 \\
Guerrillero heroico & & 0 &	\textbf{100}	& &  {none} &  {revolution} & & 0 & \textbf{57} \\
Assassination of Inejiro Asanuma & & 14 & 57 & &  {war} &  {none} & &	0 &	0 \\
Burning monk & & 0 & 14 & &	 none &  none & &	0 &	14 \\
Saigon execution & & \textbf{86}	& 14 & &  war &  war & & 0 &	14 \\
A man on the moon & & \textbf{100} &	\textbf{100} & &	 triumph	&  triumph & & \textbf{86} & \textbf{71} \\
Kent State shootings & & 57	& 57 & &  protest &  protest & & 14 &	14 \\
Accidental napalm & & 0 & 71 & &  {suffering} &  {war} & & 14	& 29 \\
Allende's last stand & & 29 & 0	& &  {triumph}	&  {war} & & 0 & 0 \\
Afghan girl & & \textbf{100}	& \textbf{100} & &  {suffering} &  {none} & &	50 & 29  \\
Tank man & & 33 & \textbf{100} & &   {war} &	 {protest} & &	0 & \textbf{71} \\
Vulture and the little girl & & \textbf{83} & 71	& &  suffering &  suffering & & 33	& 29 \\
Survivor of a Hutu death camp & & 57 & \textbf{86} & &  suffering &  suffering	& & 0 &	0 \\
Falling man & & 71	& \textbf{86} & &  terror &  terror & & 29 & 43 \\
Hijacked airplane & & 14 & \textbf{100} & &  {triumph} &  {terror} & &	14 & \textbf{57} \\
Abu Graib prisoner & &  \textbf{86} & 71 & &  {none} &  {suffering} & &	0 & 29 \\
Situation room & & 17 & 57 & &  none &  none & &	0 &	14 \\
Alan Kurdi & & 29 &	\textbf{100} & &	 suffering &	 suffering & & 0 & 33  \\
\midrule
Total & & 47 & 65 & &  suffering	&  suffering	& & 13 & 26\\
\bottomrule
\end{tabular}}
\end{table}

\subsection{Comparison survey}
\label{sec:comparison-survey}
In the second survey, participants are asked to compare the generated image side-by-side with the  iconic photograph. The goal is to assess differences in aesthetics and the depiction of the main subjects between the two images. The questions and their answers are:

\subsubsection*{Aesthetics}

\textit{Questions.} Do the two images share the same \texttt{<concept>}?

where \texttt{<concept>} is either \textit{composition}, \textit{emotional impact}, or\textit{ colour palette}.

\textit{Answer.} For each \texttt{<concept>}, participants are asked to rate in a Likert scale from $1$ (totally different) to $5$ (the same) the \texttt{<concept>} similarity  between the two images.

\subsubsection*{Demographics}

\textit{Question.} Do the main subjects depicted in the two images share \texttt{<attribute>}? 

where \texttt{<attribute>} is either \textit{gender}, \textit{age range}, \textit{ethnicity}, \textit{socioeconomic status}, \textit{period in which they were alive}, or \textit{facial expression}.

\textit{Answer.} For each \texttt{<attribute>}, participants are asked to rate in a Likert scale from $1$ (totally different) to $5$ (the same) the \texttt{<attribute>} similarity between the main subjects of the two images.

\begin{figure}
    \centering
    \includegraphics[width=0.8\linewidth]{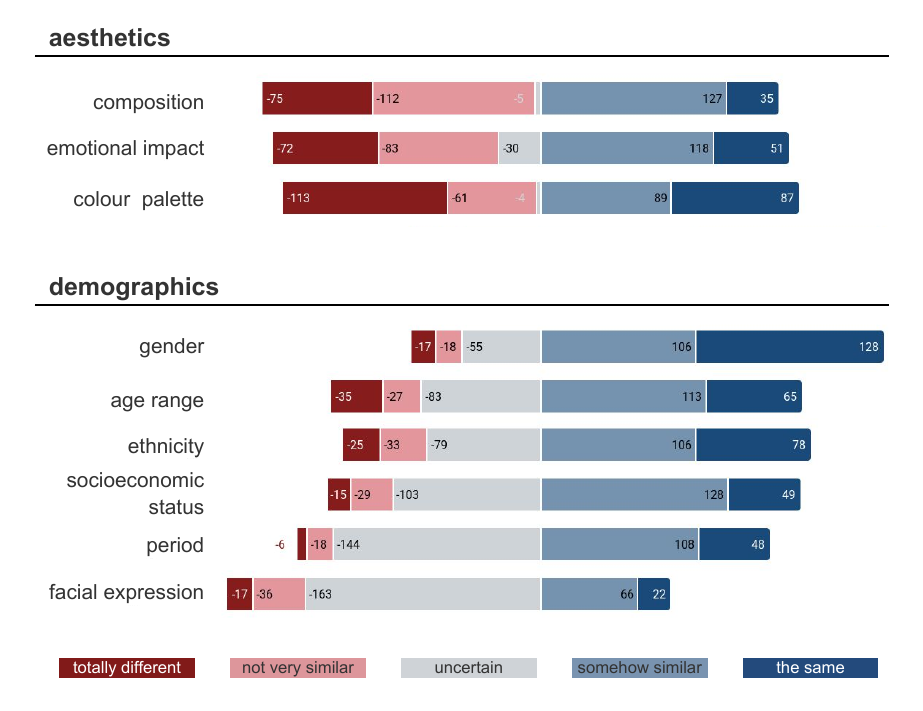}
    \caption{\textbf{Comparison survey results}. Top: results for each of the three aesthetics categories. Bottom: results for the six demographics categories.}
    \label{fig:user-study-2}
\end{figure}

\paragraph{Results}

\begin{figure}
    \centering
    \includegraphics[width=\linewidth]{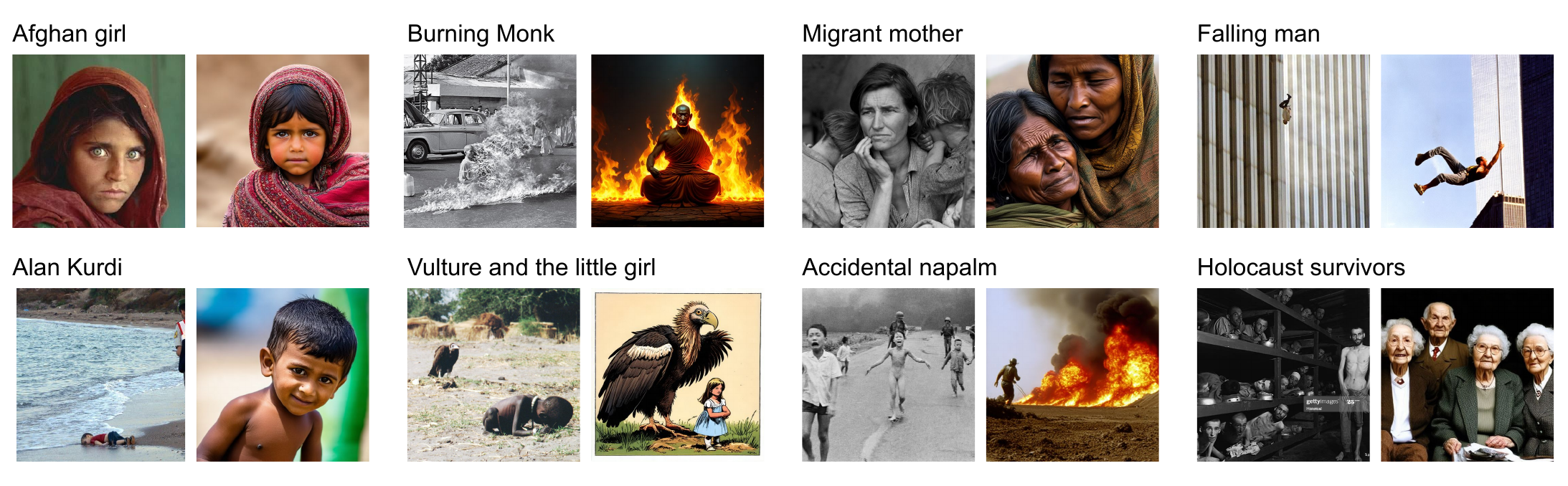}
    \caption{\textbf{Comparative examples}. Examples referenced in Section \ref{sec:comparison-survey}. For each pair, the iconic image is on the left and generated image on the right.}
    \label{fig:comparison}
\end{figure}

Results are shown in Figure \ref{fig:user-study-2} and referenced examples are provided in Figure \ref{fig:comparison}. The  aesthetics scores are polarized, indicating that while some generated images are aesthetically similar to the original iconic photographs, others differ significantly. For example, generations of \textit{Afghan girl} closely mimic the original photograph, although consistently fail to reproduce her distinctive and famed green eyes. In contrast, generations of \textit{Burning monk} show \textit{totally different} scores in the three categories: they are depicted in colour rather than black and white, the main subject is depicted alone in an ambiguous setting rather than in front of a crowd, and the fire seems to be around him rather than consuming him.

Demographic attributes have higher rates. Among them, gender shows the highest degree of fidelity to the original images. This is likely because prompts often explicitly specify the gender of the subjects, as in \textit{Afghan girl}, \textit{Migrant mother}, and \textit{Falling man}, allowing the model to accurately replicate it. When gender is not explicitly stated in the prompt, the model defaults to generate male figures, which, in many cases, aligns with the original main subject. 

For ethnicity, mismatches seem indicative of strong underlying biases, such as in the cases of images generated from \textit{Alan Kurdi}, \textit{Vulture and the little girl}, and \textit{Migrant mother}. For example, the term ``migrant" consistently fails to generate white subjects. Another notable attribute is the divergencies in the generated socioeconomic status. In \textit{Accidental napalm} the main subject is changed from a poor girl to a soldier, while generations for \textit{Holocaust survivors} represent elderly people in modern coats, indicating a recency bias and a preference for contemporary depictions over historical ones. As for facial expression, the results show the largest uncertainty, as many images (both generated and iconic) do not feature clearly visible faces.

\section{Discussion}

\subsection{Results}
Within this work we have studied the representation and contextualization of iconic images in AI image generators through a three-part analysis, comprising of a data attribution study, computational similarity analyses, and a socio-historical user study. Each of the three parts focused on a complementary aspect of how image generators engage with iconic images. From the data attribution study it became apparent that, even when prompting with the name (or a description) of the iconic image, the attribution methods struggled to identify any visual influence of the iconic image on the generated image. Only in the case of the \textit{Raising the flag on Iwo Jima} did we observe consistent strong influence, and for \textit{Guerrillero heroico} we observed strong influence only for the descriptive prompt that included `Che Guevara'. These results hint at the importance of using specific keywords, rather than appropriate semantics, to describe the iconic image.

In the computational similarity analyses we study this further by exploring the \texttt{icon2group} and \texttt{ingroup} similarities of the generated images compared against Google Search results and the underlying LAION training dataset. Here, we consistently observe that, regardless of the prompt type, the Google Search results prominently feature the iconic image, whereas the generated image benefits from more descriptive prompts. This is somewhat surprising, as searching the LAION dataset by the iconic image's name does result in many images similar to the icon, thereby pointing to effects of the training process. In terms of \texttt{ingroup} similarity there is little variation among the generated images, which may be a reflection of the guidance scale parameter used, but it similarly also reflects that the image generator was able to infer a consistent visual interpretation for the prompt.

As throughout the research process we (as researchers) have repeatedly engaged with the generated images, we have become familiar with the tendencies of the generators and how these connect to the prompts. To study how well unfamiliar observers could connect the generated images to the iconic images we performed a user study. From the user study, we observe that, regardless of the prompt, it is challenging to generate images that are recognizable with respect to the original icon, nonetheless, the participants were more often able to recognize the correct historical context, which hints at the generated images containing aspects of the icon. When visually comparing to the icon, we find significant differences in terms of aesthetics, and clear biases related to ethnicity, recency, and socioeconomic status. The user study aligns with our computational findings in that there are a few icons which seem to be well-reproduced, but for the majority of iconic images this is not the case.

Overall, and despite wide representation of the chosen iconic images on the web and in the training data, our results reveal that iconic images are not strongly represented in AI image generators. Whilst in the case of unique prompts associated with the iconic image certain effects were observed, systemic biases within AI image generators (e.g., prompts with `migrant' always return a person of color) overwrite any potential iconicity bias. Given the widespread presence of the iconic images in the data we do believe this to be an artifact of the data processing and training processes.

\subsection{Limitations}

Given the scope of this paper, which focused on the influence of iconic images on newly generated images, there are certain aspects we did not cover. Notably, through our analysis we have not studied whether the generated images could be considered iconic. Whilst we believe this may be an interesting avenue for future research, we recognize that aspects of iconicity that require widespread circulation and repeated distribution cannot be considered. Hence, we focused on the aspect of iconicity that is best studied for generated images, their representation of historical events by means of studying existing iconic images. 

Similarly, our work centered on distinct iconic images from events in the last 100 years. Whilst we believe that both older distinct and generic iconic images would be highly interesting to study, they may require a different methodology. Typically, older images concern a different modality than photography, requiring adaptions for domain gaps that may emerge from this modality difference. Whereas generic iconic images are more concerned with particular semantics rather than keywords, which would require a different prompting strategy. Moreover, we would hypothesize that AI image generators may have developed their own vocabulary of generic iconic images, i.e., recurring visual concepts and metaphors that are drawn upon, which presents an interesting avenue for future research.

The iconic images studied in this paper only present one view on which images may be considered iconic, where this view predominantly comes from a Western/US centric perspective. Whilst this does not represent the diversity of global visual cultures, it does align with the biases found in AI models and AI training datasets, which have a similar Western centric focus. Moreover, our results do point at, even within this set of iconic images, a bias towards the images about events associated with Western countries, reinforcing notions about biases in AI. To enable future research on global iconic images it is necessary for AI models to be trained on more global datasets.
\section{Conclusion}

We have sought to discover whether visual generative AI models have an iconicity bias, i.e., if iconic images have a strong influence on generated images, akin to their influence on human visual communication. To determine this, we performed a three-part analysis that studied how well visual generative AI models can reproduce iconic images for prompts with different levels of descriptiveness. Our findings show that iconic images, despite their widespread availability on the web and in the training data, are not strongly represented within visual generative AI models. Only few examples with little linguistic and visual variation can be consistently reproduced.

Moreover, the output of visual generative AI models is heavily influenced by concepts with a strong bias component, such as the word `migrant' always producing a person with a dark skin tone, even when prompted for the `migrant mother' iconic image. Similarly, prompts containing `tank' or `soldier' predominantly return images with cartoon or video game aesthetics, thereby steering the resulting image away from the iconic image described in the prompt. As such, it is clear that despite the historical and cultural importance of these images they do not have a special status for visual generative AI models.  The acontextual treatment of images during training, which disregards their importance leads to a frequency biased misrepresentation. Alternative data cleaning and training strategies that take into account the context in which images are created and used are necessary to overcome the issues uncovered in this paper.

\section*{Acknowledgments}
We thank Tim Alpherts, Selina Khan, Patrick Ramos, Ryan Ramos, Tong Xiang, Shuai Wang, Lu Wei, and Yankun Wu for participating in the user study. We also would like to thank the reviewers and attendees of the 2025 Conference on Ethics and Aesthetics of Artificial Images (EA-AI 2025) for their valuable input and feedback.

This work was partly supported by JSPS KAKENHI No. JP22K12091.

{\small
\bibliographystyle{apalike}
\bibliography{refs}
}

\end{document}